\documentstyle[PASJadd,epsf]{PASJ95}

\newcommand{\vol}{}
\newcommand{\no}{}
\newcommand{\lett}{}
\newcommand{\spage}{1}
\newcommand{\rdate}{2000 February 23}
\newcommand{\adate}{2000 March 7}
\begin{document}
\setcounter{page}{1}
\title{Discovery of a New Deeply Eclipsing SU\,UMa-Type Dwarf Nova, 
	IY\,UMa (= TmzV85)}
\author{Makoto {\sc Uemura}, Taichi {\sc Kato}, Katsura {\sc Matsumoto}\\
{\it Department of Astronomy, Faculty of Science, Kyoto University, Sakyou-ku, Kyoto 606-8502}\\
{\it E-mail (MU): uemura@kusastro.kyoto-u.ac.jp}\\
Kesao {\sc Takamizawa}\\
{\it Oohinata 65-1,  Saku-machi, Nagano, 384-0502}\\
Patrick {\sc Schmeer}\\
{\it Bischmisheim, Am Probstbaum 10, 66132 Saarbr\"{u}cken, Germany}\\
Lasse Teist {\sc Jensen}\\
{\it Center for Backyard Astrophysics (Denmark), Sndervej 38, DK-8350 Hundslund, Denmark}\\
Tonny {\sc Vanmunster}\\
{\it Center for Backyard Astrophysics (Belgium), Walhostraat 1A, B-3401 Landen, Belgium}\\
Rudolf {\sc Nov\'{a}k}\\
{\it Nicholas Copernicus Observatory, Krav\'{i} hora 2, Brno 616 00, Czech Republic}\\
Brian {\sc Martin}\\
{\it Center for Backyard Astrophysics (Alberta), The King's University College, 9125-50th Street,}\\{\it Edmonton, Alberta, Canada}\\
Jochen {\sc Pietz}\\
{\it Rostocker Str. 62, 50374 Erftstadt, Germany}\\
Denis {\sc Buczynski}\\
{\it Conder Brow Observatory, Littlefell Lane, Lancaster LA1 IXD England}\\
Timo {\sc Kinnunen}\\
{\it Sinirinnantie 16, SF-02660 Espoo, Finland}\\
Marko {\sc Moilanen}, Arto {\sc Oksanen}\\
{\it Nyrola Observatory, Jyvaskylan Sirius ry, Kyllikinkatu 1, FIN-40100 Jyvaskyla, Finland}\\
Lewis M. {\sc Cook}\\
{\it Center for Backyard Astrophysics (California), 1730 Helix Ct. Concord, California 94518, USA}\\
Tsutomu {\sc Watanabe}\\
{\it VSOLJ, 117 Shirao dormitory, 1414 Oonakazato, Shizuoka, 418-0044}\\
Hiroyuki {\sc Maehara}\\
{\it VSOLJ, Namiki 1-13-4, Kawaguchi, Saitama, 332-0034}\\
Hiroshi {\sc Itoh}\\
{\it VSOLJ, Nishiteragata-cho 1001-105, Hachioji, Tokyo, 192-0153}}
\abst{We discovered a new deeply eclipsing SU\,UMa-type dwarf nova, 
IY\,UMa, which experienced a superoutburst in 2000 January.  
Our monitoring revealed two distinct outbursts, which suggest a superoutburst 
interval of $\sim$ 800 d, or its half, and an outburst amplitude of 5.4 mag.  
From time-series photometry during the superoutburst, we determined 
a superhump and orbital period of 0.07588 d and 0.0739132 d, respectively.}
\kword{accretion disks --- binaries: eclipsing --- stars: cataclysmic 
variables --- stars: individual (IY UMa)}
\maketitle
\thispagestyle{headings}
\section{Introduction}
Dwarf novae are cataclysmic binary stars which exhibit 
repetitive outbursts of several magnitudes.  They contain 
a Roche-lobe-filling cool dwarf star that loses mass through 
the inner Lagrangian point, and a white-dwarf star accreting 
it (Warner 1995).  The SU\,UMa stars form a sub-class of 
dwarf novae,   showing two types of outburst, namely, 
a short ``normal'' outburst and a long ``superoutburst''.  
According to theories for the superoutburst 
mechanism (e.g.\,Osaki 1996), after the accretion disk 
grows over a critical radius it becomes tidally 
unstable due to a gravitational interaction with the secondary.  
In this model the precession of an eccentric disk can explain 
the ``superhump'' modulation present in the superoutburst.  

Eclipsing systems provide a unique opportunity to reconstruct 
the brightness distribution of an accretion disk from the 
observed integrated light (Horne 1985; Baptista, Steiner 1991, 1993).  
There are only five known SU\,UMa stars 
which exhibit deep eclipses, indicating occultation of the 
accretion disk and the white dwarf by the secondary star.  
Of these systems, HT\,Cas (Zhang et 
al.\,1986), OY\,Car (Krzeminski, Vogt 1985), and Z\,Cha 
(Bailey 1979) have long been studied; these limited 
samples have historically provided almost all of our knowledge 
concerning the spatial structure and time-evolution of accretion disks 
in SU\,UMa stars.  Although two more eclipsing SU UMa stars, 
DV UMa (Nogami et al., in preparation) and V2051 Oph 
(Kiyota, Kato 1998), 
have recently been discussed, the low frequency of superoutbursts 
and the small number of the known eclipsing systems still make it 
difficult to directly clarify the eccentric disk, itself, and its 
evolution with time by an observational approach.

In this letter we report on the discovery of a new deeply 
eclipsing northern SU\,UMa-type dwarf nova, IY\,UMa (= TmzV85), 
along with the results of our photometric monitoring and time-resolved 
photometry.  A more detailed analysis of the eclipses, 
including the time-evolution of the accretion disk during this 
superoutburst and the subsequent rapid fading phase, will be presented 
in a separate paper.

\section{Discovery and Observations}
Takamizawa (1998) discovered a new variable star which had 
been brightening at a photographic magnitude of 13.0 
on 1997 November, 9.751 (UT) more fainter than 14.9 mag on 
November 1.753 (UT).  He reported this star as TmzV85 to the Variable 
Star network (VSNET, {\tt http://www.kusastro.kyoto-u.ac.jp/
vsnet}) along with his comment that this is a potential 
dwarf nova based on the lack of detections on other films 
and the relatively blue color of the corresponding USNO star.  
Following this report, visual and CCD monitoring was 
conducted, yielding negative results until 2000 January 13.509, 
when Schmeer (2000) detected a second brightening at an unfiltered 
CCD magnitude of 14.0.  We started CCD time-series 
observations, which immediately revealed the presence of 
superhumps and deep eclipses, establishing that TmzV85 is a new 
deeply eclipsing SU\,UMa-type dwarf nova (Uemura et al. 2000).  
This object has subsequently been given the designation 
IY\,UMa (Samus 2000).  This is the first case in which 
both superhumps and eclipses were discovered simultaneously.  
The position of IY\,UMa, derived by H. Yamaoka (in private 
communication), is R.A. = 10h 43m 56s.87, Decl. = +58$^\circ$ 
07$^\prime$ 32$^{\prime \prime}$.5 (equinox 2000.0).  
Figure 1 gives the finding chart of IY\,UMa. 
\begin{figure}
\centerline{
\epsfysize=2.5cm
\epsfbox{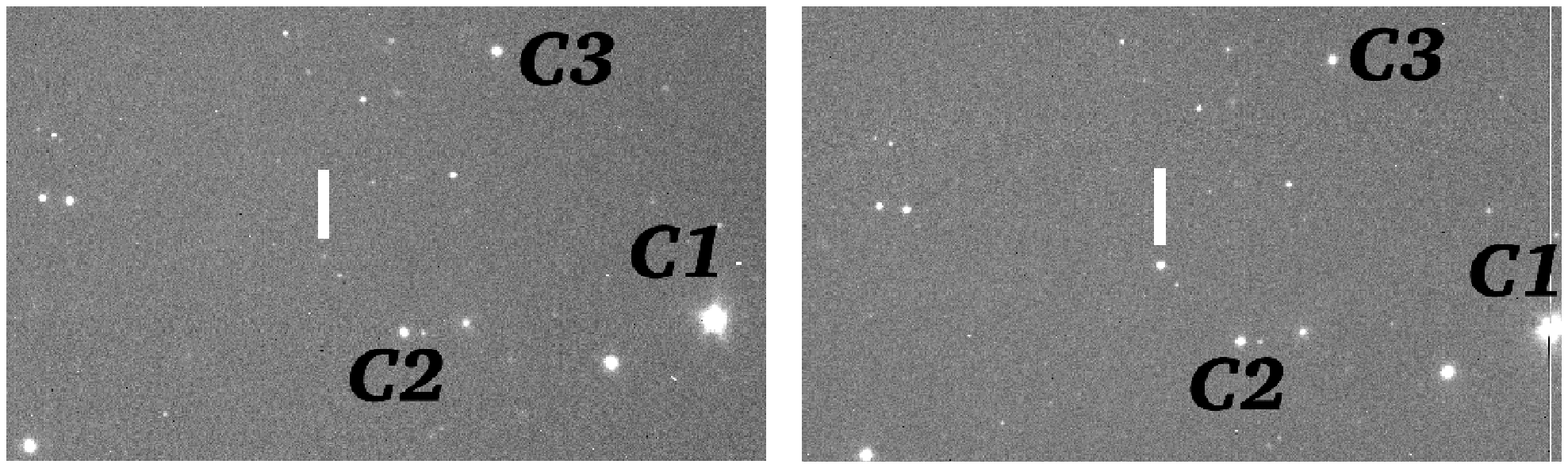}}
\end{figure}
\begin{fv}{1}
{0pc}
{CCD image of IY\,UMa observed at Iowa Robotic Observatory.  The position 
of IY\,UMa is indicated.  The differential photometry was performed with the 
comparison star C1, and its consistency was checked using stars C2 and C3.  
The left and right panels show IY\,UMa in quiescence (JD 2451552) and 
outburst (JD 2451556), respectively.}
\end{fv}
\begin{table}[t]
Table~1. \hspace{4pt}Equipment for time-series photometry.\\
\bigskip
\small
\begin{tabular}{cccc}
\hline \hline
Observer & Telescope & CCD & $T_{\rm exp}$ (s) \\ \hline 
Kyoto team & 25-cm SC & ST-7 & 30\\
L. T. Jensen & 25-cm SC & ST-6 & 60\\ 
B. Martin & 31-cm N & CB245 & 200\\ 
R. Nov\'{a}k & 40-cm N & ST-7 & 50 \\
D. Buczynski & 33-cm N & SXL8 & 45 \\ 
Nyrola team & 40-cm SC & ST-7E & 120/60 \\
J. Pietz & 20-cm SC & ST6-B & 90/60\\ 
T. Vanmunster & 35-cm SC & ST-7 & 60 \\ 
L. Cook & 44-cm N & CB245 & 16 \\ \hline
\end{tabular}
\footnotesize
\vspace{6pt}\par\noindent
SC = Schmidt-Cassegrain telescope\\
\noindent
N = Newtonian telescope
\end{table}

A description of the equipment of CCD time-series photometry 
is given in table 1.  After correcting for the standard de-biasing 
and flat fielding, we processed 
object frames with the PSF and aperture photometry packages.  
We performed differential photometry relative to the comparison star, C1, 
shown in figure 1, whose constancy was confirmed by check stars, C2 and C3. 

\section{Results}
The first outburst occurred in 1997, which was recorded as two 
photographic magnitudes of 13.0 (November 9.751) and 13.4 
(November 9.756); the second outburst was observed at 
14.0 mag on 2000 January 13.509.  The first outburst was 
most certainly also a superoutburst, because the observed 
maximum was brighter than that of the second outburst.  
This means the amplitude of a superoutburst of 5.4 mag, 
determined by the minimum magnitude of 18.4 (Henden 2000).  
No other outburst over 15.3 mag has been recorded since 
1994 November 11 when K.\,Takamizawa took his oldest photographic 
image of the field of IY\,UMa, although this object has been 
relatively closely monitored, particularly since 1999. 
Because we cannot exclude the possibility of overlooking another superoutburst 
before 1999, we suggest the typical time interval between two 
subsequent superoutbursts, called ``supercycle'', is $\sim$800 d, 
or its half.  

The light curve of the outburst in 2000 January is given in figure 2.  
The abscissa and ordinate denote time in heliocentric julian 
date and unfiltered CCD or visual magnitude, respectively.
The points and their errorbars denote the nightly averaged 
outside-eclipse magnitudes of CCD time-series photometry and their standard 
error, respectively.  The open circles denote the magnitude 
by CCD monitorings.  The last recorded magnitude 
before the outburst of 17.6 mag on January 8.463 and the 
discovery date of the outburst suggest that the duration of this 
superoutburst was between 12 and 18 days.  

Figure 3 provides the light curve on HJD 2451561.9 -- 
2451562.7, representative of the intermediate stage of the 
superoutburst (upper panel) and HJD 2451567.1 -- 2451567.9, 
representative of the advanced stage (lower panel), which clearly 
show the superhumps and the deep eclipses.    
The superhump amplitude of about 
0.5 mag in the intermediate stage becomes smaller at the late stage, 
about 0.3 mag.  On the other hand, the eclipses become 
deeper with time; the typical depth of the eclipse are about 1.3 mag 
in the upper panel and about 1.8 mag in the lower panel, 
which suggests that the brightness of the outer part 
of the disk gradually fades and/or the disk, itself, shrinks 
during the advanced stage.  The profile of the eclipse is quite asymmetric, 
suggesting the presence of a strongly asymmetric disk.  

We first determined the ephemeris of the eclipse center, $T$ (in HJD):

\begin{math}
T = 2451561.24546 (\pm 0.00011) \\
\hspace{3cm}+ 0.0739132 (\pm 0.0000018)\times E \; ,
\end{math}

\noindent
where $E$ is a cycle number.  To determine the superhump period,
we rejected observations within phase 0.10 of the eclipse center 
from all of the data obtained during the  outburst 
(HJD 2451561.2098 -- HJD 2451569.5865).  
After removing a linear trend of the decline, we performed a period 
analysis using the Phase Dispersion Minimization (PDM) method 
(Stellingwerf 1978), which indicates $0.07588\pm 0.0000113$ d 
as the best estimated superhump period.  
The superhump excess, $\varepsilon$, defined by
$\varepsilon = (P_{\rm sh} - P_{\rm orb})/ P_{\rm orb}$
, where $P_{\rm orb}$ and $P_{\rm sh}$ denote the orbital 
and superhump period, respectively, is consequently 
calculated as 0.027.  
\begin{figure}
\centerline{
\epsfysize=5cm
\epsfbox{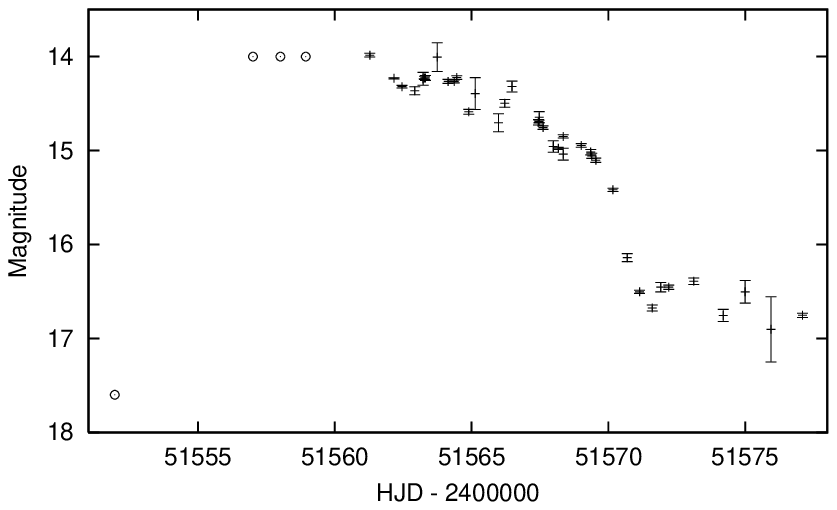}}
\end{figure}
\begin{fv}{2}
{0pc}
{Superoutburst light curve of IY\,UMa.  The outburst duration and 
the decline rate are typical for SU\,UMa stars.  The long tail after 
the outburst is characteristic for this outburst of IY\,UMa.}
\end{fv}
\section{Discussion and Summary}
We have derived some of the physical parameters of the newly 
discovered SU\,UMa-type dwarf nova with a deep eclipse, IY\,UMa, 
and summarize them in table 2 along with those of the other 
eclipsing SU\,UMa stars.  As shown in this table, the orbital 
period of IY\,UMa is quite similar to those of Z\,Cha and HT\,Cas. 
Because the normal outburst of IY\,UMa has historically not been detected, 
continuous observations are essential to determine the 
frequency of a normal outburst and the part of accretion disk 
where the disk-instability begins.  

IY\,UMa is potentially the most valuable northern SU\,UMa-type dwarf 
nova with deep eclipses, the bright quiescence magnitude, and 
relatively frequent superoutbursts compared to HT Cas, which has 
not been observed to undergo a superoutburst since 1985.  
The star thus provides a unique opportunity for SUBARU 
telescope to study the structure of accretion disks of SU\,UMa-type 
dwarf novae in quiescence through eclipse-timing spectroscopic 
observations.\par

\vspace{1pt}\par
We are pleased to acknowledge comments by D. Nogami, which 
lead to several improvements in this paper.  
This research has been supported in part by a Grant-in-Aid for
Scientific Research (10740095) of the Japanese Ministry of Education,
Science, Sports, and Culture.  KM has been financially supported as a
Research Fellow for Young Scientists by the Japan Society for the Promotion
of Science.  PS's observations were made with the Iowa 
Robotic Observatory, and he wishes to thank Robert Mutel and his
students.

\onecolumn
\begin{figure}
\centerline{
\epsfbox{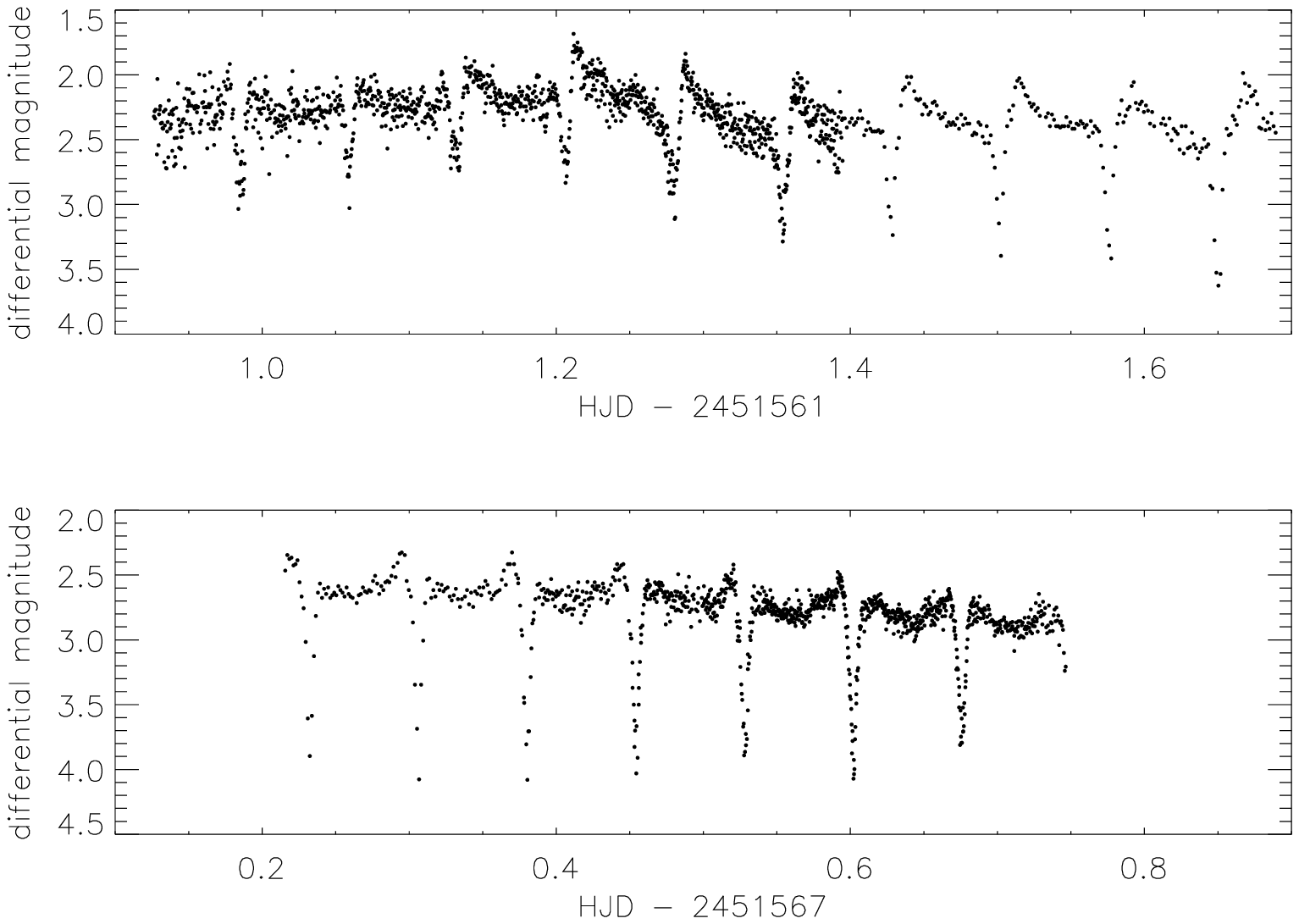}}
\end{figure}
\begin{fv}{3}
{0pc}
{Eclipses and superhumps.  Upper panel: the light curve 
on HJD 2451561 (intermediate phase in superoutburst).  Lower panel: 
on HJD 2451567 (late phase in superoutburst).  The eclipses become 
deeper with time.}
\end{fv}
\begin{table}[t]
\begin{center}
Table~2. \hspace{4pt}Physical parameters of the known eclipsing SU UMa stars.\\
\bigskip
\begin{tabular}{cccccc}
\hline \hline

Object Name & $V_{\rm quies}$ & $V_{\rm super}$ & $T_{\rm super}$ & 
$P_{\rm orb}$ & $P_{\rm sh}$\\ \hline

Z Cha & 15.3 & 11.9 & 287 & 0.074499 & 0.07740 \\

OY Car & 15.3 & 11.4 & 318 & 0.063121 & 0.064544 \\

V2051 Oph & 15.0 & 11.7$^*$ & 430$^*$ & 0.062428 & 0.06423$^\dag$ \\ 
 
HT Cas & 16.4 & 11.9$^\ddag$ & & 0.073647 & 0.076077 \\

DV UMa & 18.6 & 14.0$^*$ & 970$^*$ & 0.08587$^\S$ & 0.08867$^\S$ \\
                                          
IY UMa & 18.4$^*$ & 13.0 & 400 or 800 & 0.073913 & 0.07588 \\ \hline
\end{tabular}
\end{center}
\begin{center}
\footnotesize
\vspace{6pt}\par\noindent
${\rm V_{\rm quies}}$: magnitude at quiescence, 
${\rm V_{\rm super}}$: maximum magnitude during superoutburst, \\
${\rm T_{\rm super}}$: supercycle, 
${\rm P_{\rm orb}}$: orbital period, and 
${\rm P_{\rm sh}}$: superhump period. \\
Data without symbols from Ritter and Kolb (1998)\\
$*$ from VSNET data, 
$\dag$ IBVS 4644, 
$\ddag$ IAU Circ. 4027, 
$\S$ Uemura et al. 2000 in preparation
\end{center}
\end{table}
\twocolumn
\section*{References}
\re
Bailey J.\ 1979, MNRAS\ 188, 681
\re
Baptista R., Steiner J. E.\ 1991, A\&A\ 249, 284
\re
Baptista R., Steiner J. E.\ 1993, A\&A\ 277, 331
\re
Downes R. A., Mateo M., Szkody P., Jenner D. C., Margon B.\ 1986, ApJ\ 301, 240
\re
Henden A.\ 2000, vsnet-alert circulation 4060 \\ (http://www.kusastro.kyoto-u.ac.jp/vsnet/Mail/vsnet-alert/msg04060.html)
\re
Horne K.\ 1985, MNRAS\ 213, 129
\re
Kiyota S., Kato T.\ 1998, Inf. Bull. Variable Stars\ 4644
\re
Krzeminski W., Vogt N.\ 1985 A\&A\ 144, 124
\re
Mattei J. A., Kinnunen T., Hurst G.\ 1985 IAUC 4027
\re 
Mennickent R. E., Matsumoto K., Arenas J.\ 1999, A\&A\ 348, 466
\re
Osaki Y.\ 1995, PASJ\ 47, 47
\re
Osaki Y.\ 1996, PASP\ 108, 39
\re
Patterson J.\ 1998, PASP\ 110, 1132
\re
Ritter H., Kolb U.\ 1998, A\&AS\ 129, 83
\re
Samus N. N.\ 2000, IAU Circ. 7353
\re
Schmeer P.\ 2000, vsnet-alert circulation 4027 \\ (http://www.kusastro.kyoto-u.ac.jp/vsnet/Mail/vsnet-alert/msg04027.html)
\re
Stellingwerf R. F.\ 1978, ApJ\ 224, 953
\re
Takamizawa K.\ 1998, vsnet-obs circulation 18078 \\ (http://www.kusastro.kyoto-u.ac.jp/vsnet/Mail/obs18000/msg00078.html)
\re
Uemura M., Kato T., Nov\'{a}k R., Jensen L. T., Takamizawa K., 
Schmeer P., Yamaoka H., Henden A.\ 2000, IAU Circ. 7349
\re
Warner B. \ 1995, Cataclysmic Variable Stars, p126--215 (Cambridge Univ. Press, Cambridge)
\re
Vanmunster T.\ 2000, in preparation
\re
Zhang E. H., Robinson E. L., Nather R. E.\ 1986, ApJ\ 305, 740

\end{document}